**Ambiguity of large scale temperature reconstructions from artificial tree growth in millennial climate simulations**

O. Bothe[1,2,*], D. Zanchettin[2]

1 Klimacampus, University of Hamburg

2 Max Planck Institute for Meteorology, Hamburg

*Abstract*

The ambiguity of temperature reconstructions is assessed using pseudo tree growth series in the virtual reality of two simulations of the climate of the last millennium. The simple, process-based Vaganov–Shashkin-Lite (VS-Lite) code calculates tree growth responses controlled by a limited number of climatic parameters. Growth limitation by different ambient climate conditions allows for possible nonlinearity and non-stationarity in the pseudo tree growth series. Statistical reconstructions of temperature are achieved from simulated tree growth for random selections of pseudo-proxy locations by simple local regression and composite plus scaling techniques to address additional ambiguities in paleoclimate reconstructions besides the known uncertainty and shortcomings of the reconstruction methods.

A systematic empirical evaluation shows that the interrelations between simulated target and reconstructed temperatures undergo strong variations with possibly pronounced misrepresentations of temperatures. Thus (i) centennial scale inter-annual correlations can be very weak; (ii) the decadal range of reconstructed temperatures may be as large as the range of the temperature variations over the considered time-period; (iii) decadal variability is under-represented in the reconstructions. The misrepresentations are in part due to the temporally varying temperature-growth relations and to an apparent lack of decadal scale variability in the simulated pseudo-growth series compared to the local temperatures.

---

* oliver.bothe [at] zmaw.de

# 1. Introduction

Uncertainty in large scale paleoclimate reconstructions is usually discussed in terms of the applied statistical reconstruction methods. The possibly non-linear interplay between different limitation factors of the observable proxy information and the possibly non-stationary proxy-climate relation under changing climate regimes provide additional, often neglected sources of uncertainties.

The definiteness and uncertainty of proxy-based paleoclimate reconstructions (especially from tree-rings) have been frequently studied by utilizing networks of pseudo-proxies in simulations of past climates (e.g. Mann and Rutherford, 2002; Christiansen, 2011; Hind et al., 2012), where simulated temperature records are degraded by statistical noise to achieve synthetic proxies. These studies on pseudo proxies generally utilize rather simple noise processes, but even complex statistical noise models (e. g. von Storch et al., 2009) do possibly not fully represent the intra-annual and low frequent variability of climatic influences on the proxy evolution (e.g. temperature and moisture limitations on annual tree growth).

While reconstructions of past climates based on biological or biogeochemical data like tree growth parameters assume a uniform relationship between climate controls and biological responses (e.g. Bradley, 2011; Moberg and Brattström, 2011; Mann et al., 2012), the proxy evolution depends, in the most simple setting and for the example of tree growth, on a combination of ambient temperature and moisture conditions (Evans et al., 2006; Anchukaitis et al., 2006). Dependencies and limitations may vary under changing climate regimes (Ohse et al., 2012) or may lead to an underestimation of strong climate signals (Mann et al., 2012). Thus, assuming a uniform relation is not necessarily valid as tree growth depends non-linearly on the climatic parameters, and growth parameters result from potentially non-stationary processes. Non-uniformity becomes another source of ambiguity for reconstructions of past climate changes besides the methodological uncertainties previously discussed in the literature

(e.g. Lee et al. 2008; Ammann et al., 2010; Christiansen, 2011; Moberg and Brattström, 2011). So, in this note we ask simply: How ambiguous are (pseudo-)reconstructions representing past temperatures under changing background climate conditions, and can we neglect this ambiguity relative to the uncertainty inherent in the presently used statistical reconstruction methods? To answer these questions, we apply a simple process-based model of tree ring growth to produce surrogate proxies in the virtual reality of two climate simulations and use simple, commonly adopted regression methods to obtain pseudo-reconstructions.

Recent process based algorithms allow to compute surrogate proxies from simulation-data including the interactions and possibly non-stationary variations in the proxy-climate-relation (e.g. Hughes and Ammann, 2009; Tolwinski-Ward, 2011a; Mann et al., 2012). One such algorithm is the "Lite" version (Tolwinski-Ward et al., 2011a,b) of the full mechanistic Vaganov-Shashkin (VS) model of tree ring growth (Evans et al., 2006, Anchukaitis et al., 2006; Mann et al., 2012). The VS-Lite-algorithm is reduced to the core-processes and a monthly temporal resolution, and it validly represents climatic controls on pseudo tree-growth (Tolwinski-Ward et al., 2011a,b). Ensembles of paleoclimate simulations further facilitate the assessment of reconstruction merit by using different estimates of past climate forcings. Here, two simulations are considered with weak and with strong solar forcing amplitude (otherwise the forcing is the same; for details see Jungclaus et al., 2010) providing a subjectively stationary climate and a climate with pronounced centennial scale cold and warm periods.

Within the virtual reality of these simulations, the VS-Lite-code is, to our knowledge for the first time, employed to identify the ambiguity of tree-ring-based large scale temperature reconstructions. The paleoclimate reconstruction challenge (Hughes and Ammann, 2009; http://hurricane.ncdc.noaa.gov/pls/paleox/f?p=503) similarly utilizes the VS-Lite code to assess reconstruction merit based on real world proxy networks. The code is thoroughly tested (Tolwinski-

Ward et al., 2011a) and compares well against real-world chronologies. In the virtual reality of the climate simulations it further allows to cheaply produce an ensemble of climate reconstructions. We refer to our results as "ambiguity", as our note provides a qualitative empirical description of the indefiniteness of the reconstructions and depicts how (in)consistent the reconstructions may be relative to the target temperature. We use two simple ordinary least square methods (i) to be comparable with previous discussions (e.g. Juckes et al., 2007; Lee et al, 2008; von Storch et al., 2009) and (ii) following the recommendations by Christiansen (2011).

Methods are shortly described in Section 2. As VS-Lite computed pseudo-growth series allow for a systematic description of the relation between growth, local and hemispheric temperature and the statistical reconstructions, Section 3 describes (i) the regional recovery of local and hemispheric temperature signals in the pseudo-growth series, (ii) the sensitivity of the reconstruction methods to restrictions on the available proxies, and (iii) the reconstruction merit, before Section 4 describes the relation between target and pseudo-reconstructions and Section 5 discusses possible reasons for the shortcomings. Short concluding remarks close this note.

## 2. Methods

VS-Lite (Tolwinski-Ward et al., 2011a) computes tree growth responses to monthly means of temperature and to monthly moisture based on sums of precipitation and a simple moisture budget. The monthly growth is the minimum of the two responses scaled by the mean monthly day-length relative to the mean day-length in the month of summer solstice. We do not consider the varying insolation over the simulation period. Current annual growth is the integration of monthly values from the previous September to the current December. Thus, subsequent years overlap (following Tolwinski-Ward et al., 2011a). An ensemble of pseudo tree growth is obtained by varying in the VS-Lite-algorithm the four growth thresholds above which the growth responses are maximum or below which no growth is possible (Tolwinski-Ward et al., 2011a). Pseudo-growth is calculated for the lower (upper) temperature and moisture thresholds of 0, 3, 6 and 8.5 (9., 15., 20.) degree Celsius and 0.01, 0.02 and 0.03 (0.1, 0.3, 0.5) v/v producing an ensemble of pseudo-growth for the 108 possible permutations.

We restrict the pseudo-proxy evaluation to the domain 30°N to 75°N, where temperature signals are included in the pseudo-growth series. Pseudo-growth is computed over the 108 parameter sets and all grid points. We include ocean grid points to produce surrogates for oceanic proxies assuming that oceanic and continental climates (and in turn pseudo proxies) may differently reflect the hemispheric target climate. Input for the growth calculation is taken from the community simulations of the climate of the last millennium performed with the Max Planck Institute for Meteorology Earth System Model (MPIESM, Jungclaus et al., 2010). Two full forcing simulations are chosen with either weak (M15) or strong (M25) solar forcing amplitude (for details see Jungclaus et al., 2010).

Simplicity governs the decision for two reconstruction methods, which have been shown to give reasonable or even superior results (Hegerl et al., 2007; von Storch, 2009; Christiansen, 2011): an indirect local regression (LOC) and an indirect composite plus scaling (CPS). For discussion of reconstruction methods see e.g. Juckes et al. (2007), Lee et al. (2008), Ammann et al. (2010), Kutzbach et al. (2011) or Moberg and Brattström (2011). The reconstruction target is the average temperature for the hemispheric domain 30°N to 75°N. Inter-annual variability is included, as year-to-year variations have to be considered in process-based understanding of past climate changes from reconstructions and simulations (e.g. Zanchettin et al., 2011).

For both methods, perfect knowledge 108-member reconstruction ensembles are computed for (i) all ocean and land grid points and (ii) for land grid points only. The main focus is on two 1000-member reconstruction ensembles basing on thirty pseudo-growth series randomly selected over all possible combinations of growth limits and (i) all ocean and land grid points or (ii) all land points. We note that a virtual tree growth series does not correspond to a real world chronology.

Temperature anomalies are computed with respect to the 1971-2000 climatology, pseudo-growth series are calibrated over the period 1850-2000, and a correlation threshold ensures inclusion of a temperature signal in selected tree growth series. We note that such preselection may lead to regional biases and may amplify the signal in the calibration period relative to the retrojection period. If no correlation threshold is used (for the CPS method), correspondence between reconstructions and target weakens, measures of merit (Cook et al., 1994) worsen and biases increase. The threshold is set to an absolute value of correlation of 0.4 for LOC and an absolute value of 0.15 to the regional land and ocean average over 30°N-75°N for CPS. For both approaches the proxies are standardized by their standard deviation over the full period and multiplied by their correlation to the calibration data (either local or hemispheric). Averaging provides a composite series for the CPS approach before performing an inverse regression (Shukla and Datta, 1985; Christiansen, 2011; Moberg and Brattström, 2011). The resultant is scaled to match the variability of the target. For the LOC approach, a local inverse regression is performed against the "local" grid point temperature, the resultant matched to the calibration period temperature variability, and then the local resultants are simply averaged.

## 3. Quality of reconstructions

### a. Assessment of regionality of recovered temperature signals

The application of process based algorithms of biological development in the virtual reality of an Earth System model allows assessing whether, as to be expected, differences arise between regions on how biological pseudo proxy series recover local and hemispheric temperatures. Correlation coefficients give a first indication to the strength of the linear relationship between simulated tree growth and temperature (local or hemispheric) for different periods. Figure 1 displays the square roots of ensemble means of products between correlation coefficients over the calibration period and the period 801 to 1849. Ocean and land points are included. The target correlations mainly reflect the correlations between annual hemispheric mean temperature and annual local grid point temperatures. Figure 1 is in parts similar to the evaluation of potential predictability by Annan and Hargreaves (2012).

The purely qualitative display emphasizes the importance of mid-latitude ocean basins along the simulated oceanic fronts and the atmospheric storm tracks and jet streams. Target and local correlations agree most for the North Atlantic (North Pacific) in M25 (M15) and emphasize the storm track entrance and exit regions. Local temperatures are well recovered over the Baltic Sea, the vicinity of the Himalaya range, the coast of the Laptev Sea and parts of North America. Target correlations highlight the eastern Tibetan Plateau, central northern Eurasia, parts of Europe between 45°N and 60°N and north-eastern North America.

Of course, the evaluation recovers only the linear relationship and likely depends on the (mathematical and software) characteristics of the applied climate simulator. Nevertheless, superposing local and target correlations highlights the importance of mid-latitudinal, coastal regions of both ocean basins. "Good" proxy locations, as identified here, are usually well covered, when comparing Figure 1 to the proxy networks of Mann et al. (2009), Cook et al. (2004), D'Arrigo et al. (2006) or Christiansen and Ljunqvist (2012), although some gaps may exist (e.g. Kamchatka Peninsula). However, the main conclusion from Figure 1 is, trivially, that more oceanic proxy-records would possibly reduce the ambiguity of estimates of past climate changes (compare Annan and Hargreaves, 2012).

**b. Sensitivity to selection criteria**

Following from this regional picture, a higher correlation threshold or, similarly, the number of selected proxies may impose regional biases. Sensitivity of the reconstructed temperatures is shortly evaluated for M25 relative to these two parameters. Figure 2 plots root mean square errors (RMSE) against correlations with the target for 100-member ensembles.

For LOC, as expected, both larger correlation threshold and more proxies increase correlations, reduce RMSEs and to some extent reduce the 90% intervals, although these improvements are rather small for the changes from 45 to 60 proxies and not overly pronounced for the threshold step from an absolute correlation value of 0.4 to 0.6. Hind et al. (2012) describe similar results in a framework for simulation-reconstruction comparisons. Increasing the number of proxies particularly improves the correlation, while more demanding correlation thresholds notably reduce the RMSE of the reconstructions.

Contrastingly, the measures do not necessarily improve for the CPS approach, if more proxies are considered and/or the thresholds increase. Independent of the three studied thresholds, a selection of 15 proxies performs worst and the ensemble means for 30 proxies are also separated from the other ensembles. For 30 proxies, however, a threshold of 0.2 performs already as good as a threshold of 0.3. The other ensembles accentuate this. For 60 proxies the threshold of 0.2 results in stronger correlations and smaller RMSE than the larger correlation criterion.

**c. Measures of Merit**

We apply common measures of merit to assess the skill of a selection of reconstructions (following Cook et al., 1994; see their references) computed for each of the simulations. These are 108-member perfect knowledge ensembles for land only and for full grid coverage, and equivalently 1000-member sets from 30 randomly selected proxies. We display only results for CPS. For this method, applying the correlation thresholds limits the number of available proxies to between 355 (430) and 707 (770) proxy locations out of possible 1164 in each parameter configuration for full coverage and for M15 (M25). These numbers reduce to 200 (247) to 390 (451) if only land proxies are considered. Numbers are similar for LOC.

Calibration merit is assessed over the period 1850 to 2000 via reduction of error coefficients (RE, Cook et al., 1994) and Pearson's R² (R2). Equivalent to the paleoclimate reconstruction challenge, reconstructions are validated over the two periods 1800 to 1849 (short, abbreviated suffix: s) and 801 to 1849 (long, l). For both periods, coefficients of efficiency (CE, Cook et al., 1994) and offsets of the fitted mean from the target mean accompany R2. Figure 3 illustrates the measures of merit for CPS for M15 (top) and M25 (bottom). Positive RE and CE indicate skill relative to the climatology.

CPS calibration measures suggest general merit for both simulations (Figure 3a,b), while calibration merit is limited for LOC (not shown). The CPS reduction of error coefficients for the calibration period (REc) are nearly always positive for all ensembles and reach up to 0.8 (Figures 3c,d). Calibration Pearson's R² coefficients (R2c, Figures 3a,b) are larger than 0.3 for both simulations. Full coverage measures are higher than for land only reconstructions.

Validation period merit is to a high percentage insufficient for both methods. CPS (Figures 3c,d) validation offset generally improves between short and long validation periods with an overall range of about 0.3 K. Validation coefficients of efficiency (CEl, CEs, Figures 3e,f) and reduction of error coefficients (REvl, REvs, not shown) differ notably between long and short periods. CPS diagnostics generally improve for the long period in M25, but worsen in M15 (results are similar for LOC, not shown). Long period validation RE (not shown) and CE diagnostics are generally worse for M15 (Figure 3e) than for M25 (Figure 3f) with only full coverage perfect knowledge coefficients consistently positive. Conversely, for the short validation period reconstruction success is comparable between both simulations or the merit is even better in M15; there, REvs are mainly positive (not shown). M25 full coverage diagnostics are better than land only coefficients, but even the latter are generally larger than 0.2 for the long period. Pearson's R² for long and short validation periods (R2vl, R2vs, Figures 3g,h) scale nearly linearly against one another for both simulations. R2vs values are generally slightly higher than R2vl for M15 (Figure 3g) but the relation is in parts reversed for M25 (Figure 3h). LOC Pearson's R² are smaller than for CPS (not shown). Full coverage increases R2vs and R2vl for M25 (Figure 3g,h for CPS).

Overall, neither calibration nor short validation period merits indicate satisfactory reconstructions. For the weak forcing simulation (M15), the co-occurrence of significant calibration period correlations and negative validation CE is only excluded for the perfect full coverage CPS reconstructions, while this behavior is pronounced for the random land only CPS and both random LOC ensembles (not shown). For the strong forcing simulation (M25), merit is well for CPS, but both random ensembles commonly give negative coefficients of efficiency for calibration period Pearson's R² as large as 0.6 (not shown).

In summarizing, the merit for land only reconstructions is notably worse than for the full coverage. While best achievable measures are similar for CPS and LOC, the scatter and the number of deficient measures are larger for LOC. Target offsets and Pearson's R² do not differ much for both simulations and the two approaches, but RE and CE are worse in the weak solar forcing simulation. This does not surprise as the denominator (Cook et al., 1994) should be larger for both measures in the strong forcing simulation with its larger variability. For both simulations, good calibration period merit does not guarantee reconstruction success.

## 4. The pseudo-reconstructions

In this section, we concentrate on the random ensembles for the land-only case. Two common features arise for both approaches and both simulations (only CPS results are shown): firstly, the targeted low temperatures are at the lower range of the reconstruction 90% range (Figure 4b,e for CPS based on land points only in M15 and M25 respectively), and, secondly, this is mainly due to an underestimation of amplitude and persistence of cold excursions (Figure 4a,d for CPS based on land points only).

A positive bias emerges, which is highlighted by the residual distribution quantiles for the reconstruction ensembles (Figure 4b,e). Whereas CPS and target distributions agree rather well for positive temperature anomalies, a very strong spread is seen for the LOC regression for the warm end of the distributions (not shown). CPS slightly underestimates temperatures in the positive tail of the distribution for M25. CPS deviations are comparable between the two simulations for the negative anomalies. Note, residual distribution lines do not align parallel, but cross for single members of the reconstruction ensembles.

Besides the general underestimation of cold excursions, the CPS approach gives good results for M25, while all other cases indicate large problems in the reconstruction task. However, the CPS margin is obviously too narrow and does not capture all strong decadal and multidecadal anomalies (Figure 4). In the case of a rather modest solar climate forcing (M15), this is also present for LOC (not shown). On decadal time-scales, the ambiguity (5-95 percentiles envelope) of temperature estimates is about 0.5K for CPS and about one Kelvin for LOC.

Moving window correlations over 151 years (Figure 4c,f) evolve approximately parallel to the moving window temperature variability (not included). There is no distinct relation to the moving mean, but lower correlation coefficients occur for warm periods (less pronounced for the calibration period warming). The median of CPS moving correlations varies around 0.65 in both simulations. Notable weak correlations are found in M15 around the year 1100 and from 1300 to 1500. For M25, especially high correlations are found in the 13$^{th}$ century with a subsequent pronounced weakly correlating period in the 14$^{th}$ century. There is no reason to assume that the variability of correlations is larger than expected from stochasticity (not shown, Wunsch et al., 1999; Gershunov et al., 2001) although one expects an identifiable effect of the forcing. Moving window correlations over running mean series (11-year, 31-year) may even become negative in multidecadal periods (not shown). The smoothed series correlations are weakest in the $10^{th}$, $12^{th}$ $14^{th}$, $17^{th}$ and recent centuries (in the $11^{th}$ and $19^{th}$ centuries) for M25 (M15) without any obvious relation to the applied forcing series.

Sampling from ocean and land points (not shown) reduces the width of the 5-95% envelopes of decadal reconstruction ambiguity, but results differ only slightly concerning residual quantiles, median reconstructions and moving correlations. The strong offsets in the $13^{th}$, $15^{th}$ and $16^{th}$ centuries ($13^{th}$, $14^{th}$, $17^{th}$) in the reconstructions for the M25 (M15) ensemble improve only to a limited extent and mainly for CPS. The misfit for cold excursions remains even for perfect-knowledge ensembles (not shown) although envelopes decrease markedly for the reconstructions and for the residual quantiles.

Summing up, simple reconstruction methods give unsatisfactory results in the virtual reality of two climate simulations for the last millennium: (i) decadal-scale ambiguity of estimates and distributional inter-annual deviations are about one Kelvin; (ii) short- and longer-term extreme excursions may not be captured by the reconstructions; (iii) target and reconstructions can be uncorrelated over extended, up to centennial periods.

## 5. Discussion of reconstruction ambiguity

The presented (pseudo-)reconstructions certainly are only as good as the applied simple approaches. Many studies discuss the effect of calibration (e.g. Shukla, 1972; Shukla and Datta, 1982; Christiansen, 2011), give general evaluations of regression-based reconstructions (e.g., Ammann et al., 2010; Kutzbach et al., 2011; Moberg and Brattströmer, 2011) or show how methods influence the reconstruction results (e.g. Juckes et al., 2007; Lee et al, 2008; von Storch et al., 2009). Keeping these in mind, we focus on differences in the statistics between the pseudo-growth and the temperature series (local and hemispheric target). We have 75000 locations and individual relations between pseudo-growth and temperature series (complying with the correlation thresholds in the 108 parameter permutations) for each simulation. We sub-sample the locations considering every third latitude and every third longitude and focus on a qualitative assessment displaying selected results.

Generally, the variations in statistical properties do not exceed those expected from stochasticity. Only qualitative differences arise, which, however, may become pronounced between calibration and reconstruction periods. A null hypothesis of stationarity of the time series is rejected for less than 10 percent of the cases in both simulations for the period 801 to 1850 by the Priestley-Subba Rao test (Priestley and Rao, 1969), and the Breusch-Pagan test (Breusch and Pagan, 1979) rejects homoscedasticity of the regression relation for less than 5 percent.

If the temperature variance remains rather constant, the regression coefficients depend on the covariance between the local growth and the temperature. Local covariance (e.g. for 151 year windows) varies in ranges of about 0.5 to 1 unit over the millennial period, but the calibration-period covariance may be in the outer ranges of millennial distributions (not shown). Similarly, temperature variance varies (within expected stochastic limits) with calibration-period variability possibly outside the central half of millennial distributions (not shown). For individual pseudo-growth series, inter-annual standard deviations are by construction one over the full period but may be notably smaller or reach about two in the calibration period (not shown). Additional differences, possibly influencing the regressions, are seen in the other moments (especially skewness) and their temporal variability between pseudo-growth and temperature distributions (not shown). Considering correlations, we only point out that, for inter-annual calibration-period correlations larger than 0.4, correlations between 31-year moving average series may not reach significance or may even be negative over the calibration period

(not shown).

The associated local regression coefficients over overlapping 151-year moving windows (not shown) scatter in a range of about 0.5 units around the respective calibration-period values. Variations are usually comparable to those for equivalent correlated auto-regressive processes (simulated after fitting an AR-model to the local temperatures), but variations can be much stronger for the AR-processes.

Described variations of and differences between statistics for the temperature and pseudo-growth data emphasize the possibility of deviations from the calibration-period regression relationship over the full millennial period. Quantiles of spectral properties are presented in Figure 5a-d to identify further differences between temperature (local and target) and pseudo-growth series. For periods longer than 20 years, the hemispheric target temperature spectrum displays power equivalent to the 95 percentile values of the local spectra in M15 (Figure 5a), but the target spectra is in the range of the local median in M25 (Figure 5b). Locally in M15, (Figure 5a), spectral median and upper 95% estimates differ not much between temperature and pseudo-growth over most periodicities, but the 5% estimates indicate notably more power for centennial temperature variations, which is again less pronounced for M25 (Figure 5b).

Notable mismatches between the targets and the reconstructions occur for both simulations in the period from 1401 to 1600 (Figure 4). Wavelets focussed on this period emphasize the deviations visible in the spectra. Full wavelets are averaged over the sub-period 1401 to 1600 and quantiles of averaged wavelet spectra are plotted for periods up to about 370 years (Figures 5c,d). In M15 (Figure 5c), the hemispheric target displays more spectral power in the bands around 100 and 200 years compared to all pseudo-growth and local temperature series. Locally in M15, the 95 percentile of power is notably larger than the hemispheric target power for multi-decadal periods, but the target displays more power than the local median. The range of local power generally encloses the target. It is particularly large for multi-decadal and multi-centennial periods in M25 (Figure 5d). The power of the target temperature in M25 is dominated by the bi-centennial band possibly due to the imprint of the strong solar variations over the considered period. Both simulations differ pronouncedly from the normalized wavelet spectra for selected periods from a corresponding unperturbed millennial-scale control-simulation (shading in Figure 5c,d); shifts occur to bi-centennial periods locally, and for the target and locally also the multi-decadal band is emphasized compared to the control in both simulations but more so in M15. This again may be interpreted as imprint of the imposed volcanic and solar forcing (Zanchettin et al., 2010,2011,2012).

The different spectral and wavelet properties hint to different long term dependencies in the temperature and the pseudo-growth series. Hurst coefficients (e.g. Meko and Graybill, 1995; Cohn and Lins, 2005; von Storch et al., 2009) should reflect these. Two calculation methods were considered, the one by Higuchi (1988) and the approximation of Whittle's estimator (Beran, 1994). Results differ between the methods with larger variations for the temperature than for the ring series. We concentrate on Higuchi's method and consider only temporal scales of 6 to 384 years for which methodical differences are to some extent reduced. In M15 (M25), local temperature H values (not shown) are mostly between about 0.3 and 0.6 (0.6 to 0.8) with respective growth series H values scattering between about 0.2 and 0.7 (0.2 and the respective local temperature H). Local temperatures generally give smaller H (especially in M25) than the target temperature (0.6 and 0.85 in M15 and M25) hinting to less persistence on the local scale. This assessment of long-range behaviour extends (for the considered frequency range) the spectral evaluation by indicating that local pseudo-growth series are less likely to display long-term memory than the local temperature series. The local persistence appears to be intermittent between long-term persistence and faster fluctuations for M15, while it is mostly long-term persistent for M25, where the stronger forcing may imprint larger long-term dependencies onto the local and hemispheric climates.

In summing up, although differences in statistics are not necessarily larger than expected from random processes and the regression models are generally homoscedastic, statistics of local temperature and pseudo-growth series display pronounced changes on centennial time-scales and also their interrelations change over time. Pseudo-growth series appear to be less likely to display low-frequency variability; additionally, notable though uncertain differences occur in the long-range dependence between pseudo-growth and temperature series. Simple regression based reconstructions usually ignore the persistence of the proxies (e.g. Annan and Hargreaves, 2012). Thus, accounting for their memory (their potential predictability) could improve the reconstructions, even if proxy persistence differs from that of the temperature record.

## 6. Concluding remarks

Discussions of uncertainty in large scale paleoclimate reconstructions from proxy data generally concentrate on the applied statistical reconstruction methods. Additional uncertainty originates from the possibly non-linear interplay between different limitation factors of the observable proxy information and from the possibly non-stationary proxy-climate relation under changing climate regimes. This is generally not accounted for. Støve et al. (2012) discuss a statistical test for non-linear interrelations between temperature and its proxies using 15 proxy-records.

In this study, we utilized the simple process based "Lite" version of the Vaganov-Shashkin (VS) model of tree ring growth (Evans et al., 2006; Anchukaitis et al., 2006; Tolwinski-Ward et al., 2011a; Mann et al., 2012) to systematically study the performance of simple regression-based reconstruction methods in the virtual reality of climate simulations and to evaluate the effect of temporal variability of climate-growth relations on the reconstruction performance. To this purpose, we considered two simulations distinguished by the utilized weak and strong solar forcing amplitudes and two reconstruction methods (local regression and composite plus scaling). An ensemble of 108 pseudo-growth series at each grid point is obtained by varying the parameter values for minimum and optimum growth in the VS-Lite code. Ensembles of reconstructions are produced from the surrogate tree growth records by the two regression methods and random selections of pseudo-growth series.

In our setting, pseudo-growth series correlate best with hemispheric temperature over the oceanic storm track regions especially over their entrance and exit regions, the eastern Tibetan Plateau, northern central Europe, central Eurasia and eastern North America. This mainly reflects the relationship of "local" grid-point temperature to the hemispheric mean.

The quality of reconstructions does not necessarily improve as measured by RMSE and correlations, if criteria for proxy selection are tightened. For specified values of the thresholds, the inclusion of oceanic surrogate proxies improves the reconstruction merit. Both reconstruction approaches are similar in their best achievable merit but insufficient merit is more frequently found for the local approach. Reduction of error coefficients and coefficients of efficiency indicate less reconstruction success in the weak solar forcing simulation compared to strong solar forcing conditions, but this can be attributed to the construction of the two measures of merit (Cook et al., 1994). In our setting, good calibration merit does not ensure good reconstruction quality in either simulation.

Ambiguity of reconstructions is largest for the presented land-only reconstructions based on the local reconstruction approach but even the smallest decadal 90% envelope (for the perfect-knowledge ensemble with complete coverage and composite plus scaling) exceeds frequently 0.5 K. While perfect-knowledge ensembles with full coverage correlate consistently high (larger than about 0.6) with the target, all random reconstruction ensembles (four for each simulation) display relatively low correlating episodes and possibly large errors in the climatology.

Concerning the ambiguous character of past climate-proxy relationships in our reconstructions, statistics generally do not imply heteroscedasticity of relations or non-stationarity of pseudo-growth and temperature series. Therefore, one can assume, that the ambiguity of our reconstruction ensembles is to a large extent attributable to the reconstruction approaches and the general uncertainty of reconstructing a large-scale climate variable from scarce proxy data. Even though variations of local temperature, pseudo-growth and their interrelations remain within the levels expected from stochasticity, such variability of relations is already going to affect the quality of the linear-regression-based reconstructions, as becomes obvious in the trends of the correlations between target and ensembles. While the results differ between the two forcing scenarios, the forcing appears to be neither a dominant factor in the pseudo-growth nor for the reconstruction merit. We particularly note that, for the calibration period, the simple regression relevant statistics covariance and variance may be outside the central half of their millennial distributions. Additionally we stress that even if our pseudo-reconstructions would consider the memory in our pseudo-growth proxies (i.e., their potential predictability), the long term and spectral characteristics differ prominently between the proxies and the target time series. The pseudo-growth and local temperature series are less likely than the hemispheric target to present low-frequent variability especially for centennial-scale variations.

In conclusion, the simple reconstruction methods give unsatisfactory results for pseudo-tree-growth series in the virtual reality of the two selected climate simulations for the last millennium. Ambiguity is not only inherent to (our) statistical reconstruction methods, for whose uncertainties one can account for by building prediction intervals. Additional ambiguity arises from the differences in the statistical properties of biogeochemical proxies, the climate variable they are meant to represent and the respective target. The ambiguity may be as large as the range of decadal variations of the target. This indefiniteness includes the possibility, that (i) extreme excursions and their duration are not captured and (ii) that target time-series and reconstructions are episodically unrelated inter-annually and on low-frequent time-scales.

*Acknowledgement*

We thank Susan Tolwinski-Ward for providing the VS-Lite code. O.B. acknowledges funding through the Cluster of Excellence ``CliSAP'', University of Hamburg, funded through the German Science Foundation (DFG). D.Z. was supported through the ENIGMA project of the Max Planck Society and from the Federal Ministry for Education and Research in Germany (BMBF) through the research program ``MiKlip'' (FKZ:01LP1158A). This work has been carried out as part of the MPI-M Integrated Project Millennium and contributes to the Cluster of Excellence CLISAP at Hamburg University. The Priestley-Subba Rao and the Breusch-Pagan tests have been implemented, respectively, by William Constantine and Donald Percival and Hothorn et al. for R, whereas Martin Mächler authored the R-routines for the Hurst coefficient calculation.

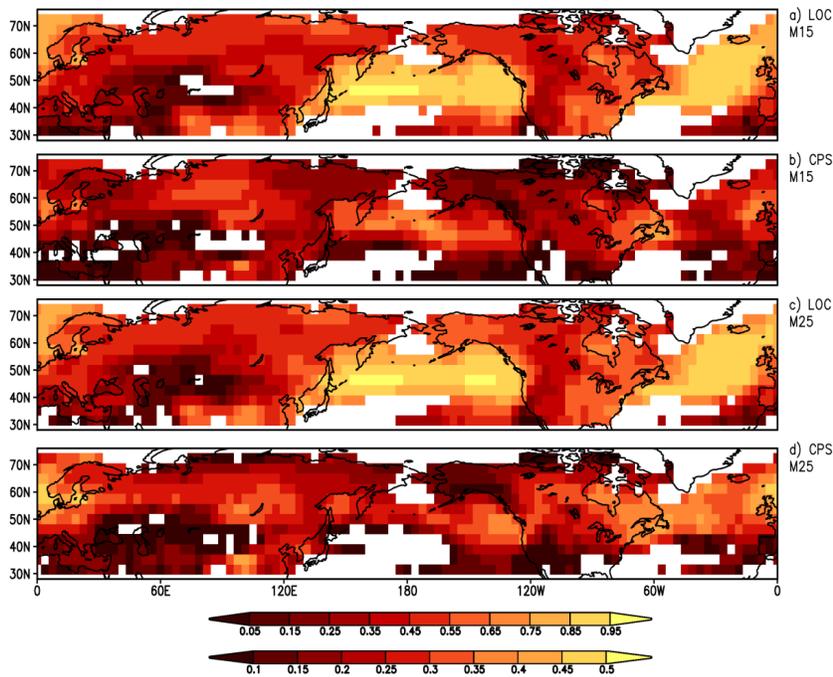

Figure 1: Squareroot of ensemble mean product of correlation coefficients for a,c the local (LOC), and b,d, the composite plus scaling (CPS) approach. Upper two panels for the weak solar full forcing simulation (M15), lower two for the strong solar full forcing simulation (M25). Upper (lower) color bar for LOC (CPS). For LOC, correlation coefficients between local temperature and growth over the calibration period and over the period 801 to 1849 were multiplied, while for CPS maps correlation coefficients were used between tree growth and temperature target for the hemispheric band 30°N-75°N for the two periods. White areas are missing values or regions where the product of correlation coefficients is negative.

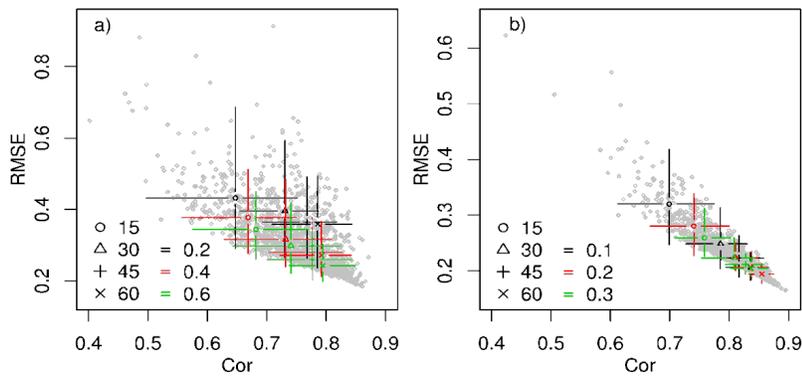

Figure 2: Sensitivity of reconstructions for the strong solar forcing simulation with respect to the number of selected proxies (symbols) and correlation thresholds (colors) for the local (left) and composite plus scaling (right) approach. Plotted are RMSE for the period 801 to 1850 against correlations with the target for this period for a 100-member ensemble. Note the different scales. Error bars are the 90 percent intervals. Dots are all 1200 members. This visualisation is not meant to evaluate both methods against one another. All grid points are included over land and ocean.

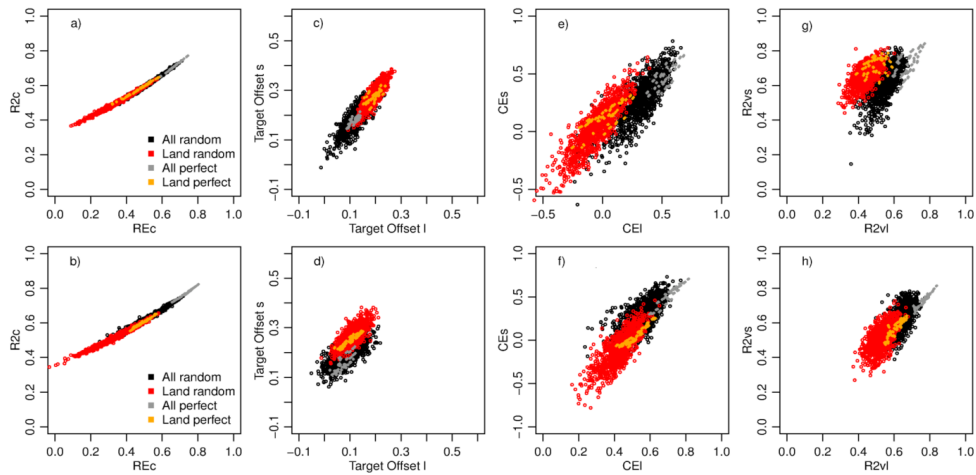

Figure 3 Composite plus scaling measures of merit for top, M15, bottom, M25: black, full coverage random selection, red, land only random, grey, perfect full coverage, orange, perfect land only: a,b) calibration R² against calibration RE, c,d). Offset of fitted mean from target mean (in Kelvin), short period versus long period, e,f) short against long validation period CE, g,h) short versus long validation R². Scales differ to some extent between x and y axes and between the approaches.

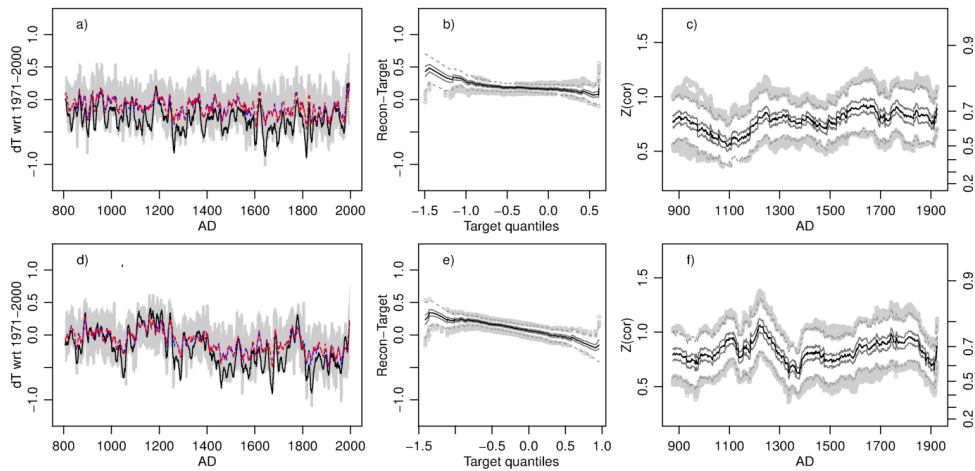

Figure 4: Random laond-only based CPS pseudo-reconstruction ensemble for, top, M15, bottom, M25: a,d) 11-year running mean target temperature in black, 5% and 95% quantiles for 11-year moving windows as grey shading and mean (blue) and median (red) over moving windows for the reconstructions; b,e) boxplot statistics of residual quantiles over 1000-member ensembles (median (black), first and third quartiles (solid grey), data within 1.5 interquartile ranges from the quartiles (grey dashed) and outlier (grey dots)); c,f) box plot statistics of Fisher's Z-Scores (left axes) of moving 151-year correlations (right axes), colors as in b,e.

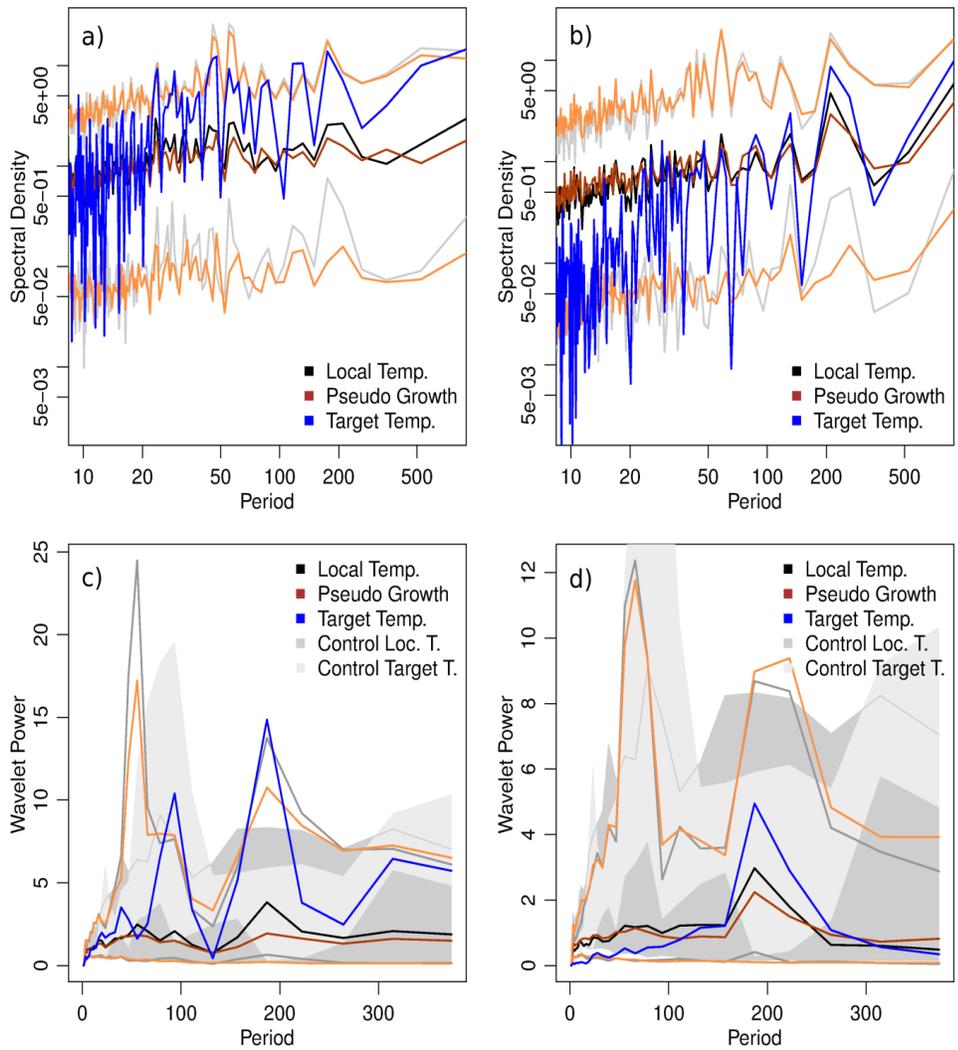

Figure 5: Left: M15, right: M25. Top: spectra for temperatures and growth series. Bottom: average wavelet power over the period 1401 to 1600. Local temperature median and 90 percent range (black and grey lines), pseudo-growth median and 90 percent range (dark and light brown) and target temperature (blue). Shadings are equivalent percentile ranges for overlapping windows of the control run for local temperature (dark grey) and the target temperature (light). Where shadings mask each other overlap, a grey line marks the local percentile range. X-axes are periods in years.